\documentstyle[prd,preprint,aps]{revtex}
\tightenlines
\begin{document}

\centerline{\Large \bf Quartic Anharmonicity in
                       Different Spatial Dimensions}

\vskip 10mm

\begin{center}
            G.V.~Efimov and G.~Ganbold\footnote{
{\it Permanent address:}
Institute of Physics and Technology, Mongolian Academy
of Sciences, \\ 210651 Ulaanbaatar, Mongolia} \\[5mm]

{\it Bogoliubov Laboratory of Theoretical Physics, \\
Joint Institute for Nuclear Research, 141980 Dubna, Russia}
\end{center}

\vskip 5mm

\begin{abstract}
A path-integral method effective beyond the perturbation expansion
approach is suggested to consider the quartic anharmonicity in
different spatial dimensions. Due to an optimal representation of
the partition function, the leading term has already taken into account
the correct strong-coupling behaviour. In the simplest cases of zero
and one dimension we have obtained reasonable results in a simple way.
Then, this technique is applied to the superrenormalizable scalar
theory $\phi^4_2$ in two dimensions. This results in an accurate
estimation of the ground-state energy that provides exact weak- and
strong-coupling behaviour already in the leading-order approximation.
The next-to-leading terms give rise in insignificant corrections.
\end{abstract}

\vskip 5mm

\section{Introduction}

The anharmonic oscillator plays an important role by providing a
theoretical laboratory for the examination of new calculational
schemes and approximation techniques (see, e.g. \cite{guil90,arte90})
in quantum mechanics. Besides, it describes more or less real self
interaction disturbing the idealized picture of the free harmonic
oscillators. The most known version is the quartic oscillator
\cite{turb89,klei90}. On the other hand, scalar fields play a
fundamental role in the unified theories of strong, electromagnetic
and weak interactions. The mechanism of symmetry rearrangement specific
for scalar fields occurs in realistic scalar field theories
\cite{linde79}. At the same time, the mathematical structure of scalar
field theory is simpler than that of the vector and spinor cases.
A native and well-known generalization of the quartic oscillator is
the model of scalar self-interaction $\phi^4$ in quantum field theory.
Particularly, the quantum mechanical quartic oscillator with a
Hamiltonian
\begin{equation}
H=\frac{1}{2}(p^2 + m^2 q^2 + g^2 q^4) \label{Hamil}
\end{equation}
can be considered \cite{bend69} a theory of a scalar field in
one-dimensional ($q(t),~t\in R^1$) Euclidean space-time described by
the Lagrangian:
\begin{equation}
L[q,\dot{q}]
={1\over2}\left[\dot{q}^2(t)-m^2q^2(t)-g^2q^{4}(t)\right] \,.
\label{Lagr}
\end{equation}
Thus, the scalar field model (\ref{Lagr}) represents an attractive,
simple but nontrivial object for research. The problem of finding the
energy spectrum $\{E_n\}$ and the proper wave functions $\{\Psi_n\}$
of the Hamiltonian can be solved by using the partition function and
path-integral formalism. Note, the corresponding Schr\"{o}dinger
equation cannot be solved explicitly for general case, except the
weak- and strong-coupling limits.

Further extension of the model (\ref{Lagr}) can be done by considering
the quartic self interaction $\phi_d^4(x)$ in higher spatial dimensions
$x\in R^d,~~d\ge 1$. Earlier investigations of the triviality problem
have shown that for dimensions $d>4$ the scalar theory of self
interaction behaved either unstable or trivial. Much interest is
attracted to the cases $d=2$ and $d=3$ demonstrating nontrivial phase
restructure.

Nowadays, the anharmonic oscillator is a system well understood in
both perturbative and non-perturbative aspects \cite{bend69,hioe78}.

Typically, the problem is treated by using various perturbative
expansion scheme. Particularly, the conventional loop expansion
\cite{jack74} gives asymptotic Rayleigh-Schr\"odinger series for
energy levels which numerically converges only for $g\ll 1$. The
loop expansion for composite operators \cite{corn74} and optimized
expansion \cite{okop87,okop93} provide better convergence for
numerical series; however, in this case the approximation becomes
worse as $g$ increases. To obtain a reasonable result, one needs
in various resummation techniques for these divergent series
\cite{bend78}. Particularly, the Heaviside transformation of the
mass parameter \cite{yama96} and a modified Laplace transformation
method \cite{mizu97} can result in good approximations to the ground
state energy, but the perturbation series depends on cut-off parameter
$x^{*}$ and converges very slowly \cite{mizu97}. The P\`ade
approximants and the Borel transform can be used to summarize the
formally diverging series \cite{graf70}. But, rigorous proofs of the
convergence can be made only in particular cases \cite{simo70} and
the convergence is nonacceptably slow.

Going beyond the perturbation formalism, a variational interpolational
method \cite{klei95} and the strong-coupling expansion \cite{weni99}
are used to estimate the energy spectrum of the AHO. Also, by applying
the renorm-group equation, one obtains a resummed perturbation series
\cite{chen96}. The renorm-group methods may improve the perturbative
expansions \cite{fras94,kuni98}. However, the agreement with the WKB
result becomes worse in the higher orders than the fourth at which the
agreement is the best. Recently, it is have shown that multiple-scale
\cite{bend96} and reductive perturbation theory \cite{bend78} can be
successfully applied to the quantum AHO to construct the asymptotic
behaviour of the wave function for large $x$. A similar method had been
proposed in \cite{gins78}. But, it is not trivial to identify the
secular terms for the  wave functions which can be made to vanish at
a renormalization point for general case.

Within the method of orthogonal polynomials it has been established
\cite{cicu96} that the solutions of the bound state equation for the
quartic anharmonic oscillator are related to a sequence where the
potential is a even polynomial of degree six (a quasi-exactly solvable
model). A regular iterative method for the ground state energy of
quartic oscillator as an analytic function of coupling constant
\cite{cicu96} is formulated. But this method introduces a PI averaging
of the complicated many-point potential of the logarithmic type that
makes it hard to obtain explicit solutions. By using an extension of
quasi-exactly solvable models \cite{ushv94} to Bose systems, one
obtains (in some particular cases) simple exact expressions for several
energy levels of an anharmonic Bose oscillator \cite{doly00} beyond the
exploit perturbation theory. A relationship between the $x^{2M}$ AHOs
and the $A_{2M-1}$ TBA systems have been conjectured \cite{dore99} by
using an alternative integral expression for the spectral determinants
of the quartic oscillator.

Therefore, it is important to build a method which is effective in the
strong-coupling. On the other hand, it has to work even in the case of
renormalizable theories (e.g., $\phi^4_2(x)$) with formally divergent
or complex functionals, unlike the variational methods. Besides, the
desired method should be relatively plain and has to catch the main
strong-coupling contribution within the leading-order approximation.
Below we represent a path-integral technique obeying these features.

In our earlier works \cite{efim89,efim92} the problem of strong
coupling regime and phase structure of the $\varphi^4_2$ quantum field
theory has been investigated. We have found that an approximate
solution of the problem can be obtained by combining the canonical
transformations of field variables and normal ordering of the creation
and annihilation operators. In the lowest approximation, the solution
obtained there is identical to the result of the variational method of
the Gaussian effective potential \cite{stev85}. Meanwhile, in contrast
with the variational method our suggested technique ensures a canonical
structure of the theory and, hence, provides calculation of
perturbation corrections over effective coupling constants which can
also be used to control an accuracy of approximation. Later on, the
method was generalized and applied to renormalizable quantum field
theories,  path integrals and  quantum mechanics of bound states
\cite{book95}.

In the present paper we demonstrate a significant modification of our
method suitable to investigate the $\phi^4$ model effectively  even for
large coupling. Below we consider the case of low dimensions, namely,
$d=0,1,2$ to avoid nonprincipal but lengthy renormalization procedures
occurring for  $d=4-\epsilon,~~~\epsilon\to 0$. In fact, an extension
of our method from $d=2$ to $d=4-\epsilon$ does not meet any principal
troubles, because only renormalization of the mass is sufficient to
remove all divergencies in both models. Therefore, upgrading the
dimensionality to $d=4-\epsilon$ will affect only some coefficients and
factors, not changing the very form of our final expressions. However,
we should note that the consideration of the theory $\phi^4_4$ requires
a serious modification of our method because one has to workout
first an effective technique to perform the  renormalization of
the coupling constant.

Our basic idea is to find a optimal representation of the partition
function (e.g., see (\ref{ZVg})) in which the main contribution
is taken into account correctly for large $g^2$ already in the lowest
approximation. For this purpose we will use the Gaussian representation
of the initial quartic interaction in the form of an integrable
quadratic one by involving an auxiliary integration as follows:
\begin{equation}
\int\!\!\delta\phi\exp\left\{ -{1\over 2}(qD_0^{-1}q)
-{g^2\over 2} q^4 \right\} \propto \int\!\! dA \exp\left\{
-{A^2\over 2}-{1\over 2} {\rm tr}\ln(1+2igqD_0)\right\} \,.
\end{equation}
Then, the inner path integral over $\phi$ can be taken out explicitly
and the remaining integral over auxiliary variable (field) $A$ can be
evaluated effectively, in particular, by isolating the total Gaussian
contribution.

\section{A Non-Gaussian Integral: $d=0$}

Below we demonstrate the basic idea of our approach on the simplest
case of a plain non-Gaussian integral.

First, we introduce a notation indicating an averaging with respect
to a normalized Gaussian measure as follows:
\begin{equation}
\left\langle (\bullet) \right\rangle_{x} =
\int\limits_{-\infty}^{\infty}\!\! {dx\over \sqrt{2\pi}}
\,e^{-x^2/2}(\bullet)\,,
\qquad   \left\langle 1 \right\rangle_{x} = 1 \,.
\end{equation}

Let us define a "partition function":
\begin{equation}
Z[g,J]=\left\langle\exp\left\{-{g^2\over 2}x^4+ixJ \right\}\right\rangle_{x}\,.
\end{equation}

Then, we can consider the zero-dimensional analogues of the Quantum
Mechanical ground-state energy and Green function as follows:
\begin{equation}
\exp\{-E(g)\}=Z[g,0] \,, \qquad \left.
G(g)=-{1\over Z[g,0]} {d^2 Z[g,J]\over dJ^2} \right\vert_{J=0}.
\label{energygreen}
\end{equation}

As is known, these integrals are not solvable exactly. According to
our approach we write
\begin{eqnarray}
Z[g,J] &\!\!=\!\!& \left\langle \int\limits_{-\infty}^{\infty}
\!\! dx\exp\left\{ -{x^2\over 2} - igx^2 A + ixJ \right\}
\right\rangle_{A} \nonumber  \\
&\!\!=\!\!& \left\langle \exp\left\{-{1\over 2}\left[ \ln\left(
1+2igA\right)+{J^2\over 1+2igA} \right] \right\} \right\rangle_{A}\,.
\label{part1}
\end{eqnarray}

To build the optimal oscillator basis, we have to eliminate the
linear $A$ terms out of the exponent as follows:
\begin{equation}
-\frac{aA}{2ig}-\frac{igA}{1+a}=0 \qquad \Longrightarrow \qquad
 g^2=a\,(1+a)\,, \qquad h={g^2\over 2}={a\,(1+a)\over 4} \,.
\end{equation}

After a shift $A\rightarrow A+a/2ig$ with a parameter $a>0$ we obtain
\begin{equation}
Z[g,J] = \left\langle \exp\left\{ -{1\over 2} \left[ \frac{aA}{ig}
-\frac{a^2}{4g^2} + \ln\left( 1+a+2igA\right) +\frac{J^2}{1+a+2igA}
\right] \right\} \right\rangle_{A} \,.
\end{equation}

First we rewrite the logarithm in the exponent as
\begin{eqnarray}
\ln \left( 1+a+2igA\right)
=\ln \left( 1+a\right) +\frac{2igA}{1+a}
+\frac{2g^2A^2}{\left( 1+a\right) ^2}+\ln_2\left(1+a+2igA\right)\,,
\end{eqnarray}
where
\begin{equation}
\ln_2\left( 1+z \right) =\ln \left( 1 + z \right) - z + {z^2\over 2}
\,, \qquad \forall \mathop{Re}z\ge -1 \,.
\end{equation}

Define a function
\begin{equation}
W(s)=-{1\over 2} \ln_2\left( 1 +i s \lambda \right) \,, \qquad
\lambda = \sqrt{{2a\over 1+2a}} < 1 \,.
\end{equation}

Then, we obtain
\begin{equation}
Z[g,J] = \exp \left\{ \frac{a^2}{8g^2} - \frac{\ln( 1+a )}{2} \right\}
\cdot\left\langle \exp \left\{ W(s) - \frac{J^2}{2(1+a)(1+is\lambda)}
\right\} \right\rangle_{s} \,.
\end{equation}

Accordingly,
\begin{eqnarray}
&& e^{-E(g)} = e^{-E_0(g)} \cdot \left\langle e^{W(s)} \right\rangle_{s} \,,
\qquad E_0(g)= -{a\over 4(1+a)}
+ {\ln\left( 1+a\right)\over 2} \,, \nonumber\\
&& G(g) = {1\over 1+a} \cdot \left\langle {1\over 1+is\lambda} e^{W(s)}
\right\rangle_{s} / \left\langle e^{W(s)} \right\rangle_{s} \,.
\label{desired}
\end{eqnarray}

Systematically expanding the exponentials in (\ref{desired}) we
estimate the "ground-state energy"
\begin{eqnarray}
\label{energy}
E_N(g)=E_0(g) + \sum\limits_{j=1}^{N} \Delta E_j(g) \,, \qquad
\Delta E_1(g) = - \left\langle W(s) \right\rangle_{s}
\end{eqnarray}
and the "Green function"
\begin{eqnarray}
\label{green}
G_N(g) &\!\!=\!\!& G_0(g)+\sum\limits_{j=1}^{N} \Delta G_j(g) \,,
\qquad G_0(g)=\frac{1}{1+a} \left\langle {1\over 1+is\lambda}
\right\rangle_{s} \,, \\
\Delta G_1(g) &\!\!=\!\!& \frac{1}{1+a} \left\{ \left\langle
{W(s)\over 1+is\lambda} \right\rangle_{s} - \left\langle
{1\over 1+is\lambda} \right\rangle_{s} \left\langle W(s)
\right\rangle_{s} \right\}  \,, \nonumber
\end{eqnarray}
where $E_0(g)$ and $G_0(g)$ are the leading-order approximations.

In Table 1 we represent our estimates due to (\ref{energy}) and
(\ref{green}) compared with the exact numerical solutions to
(\ref{energygreen}). We see that our approach describes well these
functions even within the zero-order approximation. The
next-to-leading corrections systematically improve the obtained
results approaching close to the exact solutions. Our technique
works well at least in zero dimension.

\section{Anharmonic Oscillator: $d=1$}

The most known example of the quartic interaction is the anharmonic
oscillator $g^2 q^4/2$ in Quantum Mechanics. The main problem is to
calculate its energy spectrum and to find the corresponding wave
functions.

Conventionally, the problem is treated by using various perturbative
methods (e.g., \cite{jack74,corn74,okop93}). However, the ground-state
energy in the naive perturbation series \cite{bend69} reads,
$E=m\sum_{n=0}^{\infty} A_{n}\cdot (g^2/m^{3})^n$, where the
coefficient $A_{n}$ grows as $A_{n}\sim (-3/2)^{n}\Gamma(n)$ for large
$n$ \cite{bend73}.  Thus, the series formally diverges for any finite
coupling. Only for sufficiently small coupling constant,
$g^2/m^{3} \sim 0.1$, the series becomes numerically useful by the
appropriate truncation. Beyond the perturbative approach,
variational \cite{klei95} and strong-coupling expansion methods
\cite{weni99} can be applied to this problem.

Our effective nonvariational approach described in previous section
can be extended to the quantum mechanical anharmonic oscillator
as follows.

Consider a normalized partition function
\begin{eqnarray*}
Z_V(g)=\int Dq~e^{-{1\over2}(qD_0^{-1}q)-{g^2\over2}(q^4)} \,,
\qquad Z_V(0)=1 \,,
\end{eqnarray*}
where $q(t)$ is the position vector and
$t\in [-T,T],~\int\! dt=V \to\infty$.

The differential operator $D_0^{-1}$ and its Green function $D$
obeying the conventional boundary conditions are:
$$
D_0^{-1}(t,s)=\left(-{d^2\over dt^2}+1\right)\delta(t-s), \qquad
D_0(t)={e^{-|t|}\over2},~~~~~~\tilde{D}(k)={1\over k^2+1} \,.
$$
For further convenience we use the unit mass $m=1$.

First, we introduce an additional integration variable (an auxiliary
trajectory) $\phi$ to convert the initial quartic interaction to an
integrable quadratic one by using the following Feynman representation
\begin{eqnarray}
Z_V(g)=\int D\phi ~e^{-{1\over2}(\phi\phi)} \int Dq
~e^{-{1\over2}(q D_0^{-1} q) -ig(q^2\phi)} = \int D\phi~e^{-{1\over 2}
(\phi\phi)-{1\over2}{\rm Tr}\ln[1+2ig\phi D_0]}    \,.
\label{ZVg}
\end{eqnarray}

After this transformation, the interactional functional reaches its
minimum distinct from the origin $\phi=0$. Therefore, a shift of the
variable may be suggested
$$
\phi\to \phi+{a\over2ig} \,.
$$

Then, one rewrite
\begin{eqnarray*}
Z_V(g)&=&\exp\left\{
{1\over 8 g^2}(aa)-{1\over2}{\rm Tr}\ln[1+a D_0]\right\} \nonumber\\
&& \cdot \int\!\! D\phi~\exp\left\{ -{1\over2}(\phi\phi)-{1\over 2ig}
(a\phi)-{1\over2}{\rm Tr}\ln[1+2ig\phi D]\right\} \,,
\end{eqnarray*}
where
$$
D={1\over 1+aD_0}D_0\,, \quad
D(t)=\int{dk\over2\pi}~{e^{ikt}\over k^2+\mu^2}
={e^{-\mu |t|}\over 2\mu} \,, \quad \mu=\sqrt{1+a}   \,.
$$

The optimal value of the shift parameter $a$ obeys the equation:
\begin{eqnarray*}
-{1\over 2ig}(a\phi)-{2ig\over2}{\rm Tr}[\phi D]=0  \quad
\Rightarrow \quad a=2g^2\int{dk\over2\pi}\cdot{1\over k^2+\mu^2} \,,
\qquad g^2=a\sqrt{1+a} \,.
\end{eqnarray*}
In particularly, $a=g^{4\over 3}$ for $g\to\infty$. By using the
optimal shift we rewrite
\begin{eqnarray*}
Z_V(g) &=&\exp\left\{
{1\over8g^2}(aa)-{1\over2}{\rm Tr}\ln[1+a D_0]\right\} \\
&& \cdot\int D\phi~\exp\left\{-{1\over2}(\phi{\cal D}^{-1}\phi)
-{1\over2}{\rm Tr}\ln_2[1+2ig\phi D]\right\} \,,
\end{eqnarray*}
where
\begin{eqnarray*}
&& {\cal D}^{-1}(t)=\delta(t)+2g^2D^2(t) \,, \qquad
\tilde{\cal D}^{-1}(k^2)=1+{2a\over k^2+4(1+a)}\,, \\
&& {\cal D}(t)=\delta(t)-{a\over\nu}~ e^{-\nu |t|}\,, \qquad
\tilde{\cal D}(k^2)=1-{2a\over k^2+\nu^2} \,, \qquad \nu=\sqrt{4+6a} \,.
\end{eqnarray*}

By introducing a normalized Gaussian measure
$$
d\Sigma_{{\cal D}} = {D\phi\over\sqrt{{\rm det}{\cal D}}}
~e^{-{1\over2}(\phi{\cal D}^{-1}\phi)} \,, \qquad
\int\!\! d\Sigma_{{\cal D}} =1
$$
we obtain
\begin{eqnarray}
\label{hehe1}
Z_V(g) &=& e^{-V(E_0(g)+\triangle E(g))}  \,,         \nonumber\\
E_0(g) &=& {1\over V} \left\{ -{(a\cdot a)\over 8g^2}
           +{1\over2}{\rm Tr}\ln[1+a D_0]
           +{1\over2}{\rm Tr}\ln[1+2g^2D^2] \right\}  \nonumber\\
&=& -{1\over 2}-{a\over 8\sqrt{1+a}}-{1\over 2}\sqrt{1+a}
    +\sqrt{1+{3\over 2}a} \,,                      \nonumber\\
\triangle E(g) &=&  -{1\over V} \ln \int d\Sigma_{{\cal D}}
\exp\left\{ {1\over 2} {\rm Tr}\ln_2 [1+2ig\phi D] \right\} \,.
\end{eqnarray}

We consider $E_0(g)$ the leading-order approximation to $E(g)$. It can
be compared (in Table 2) with our earlier result obtained within the
oscillator approximation method \cite{book95}:
\begin{equation}
E_{osc}(g)= -{1\over 2} + \mathop{min}_{\xi,\rho} \left\{
\frac{\xi}{8\rho} \frac{\Gamma(2-\rho)}{\Gamma(1+\rho)}
+\frac{1}{6\xi}\frac{\Gamma(1+3\rho)}{\Gamma(1+\rho)}
+\frac{a\sqrt{1+a}}{10\xi^2}\frac{\Gamma(1+5\rho)}
{\Gamma(1+\rho)}\right\}
\end{equation}
and with a direct numerical calculation $E_{num}(g)$ (see, e.g.
\cite{hioe78}).

In the weak-coupling limit ($g \ll 1$) we have:
\begin{eqnarray}
E_0(g) &\!\!=\!\!& {3\over 8} g^2-{11\over 32} g^4+O(g^6)\,,
\qquad E_{osc}(g)={3\over 8} g^2-{9\pi^2-70\over 16(\pi^2-8)} g^4
+O(g^6) \,, \nonumber\\
E(g)   &\!\!=\!\!& {3\over 8} g^2-{21\over 32} g^4+O(g^6) \,.
\label{weak3}
\end{eqnarray}

For $g\to\infty$ one obtains
\begin{eqnarray*}
E_0(g) &\!\!\to\!\!& g^{2\over 3}\left(\sqrt{{3\over 2}}-{5\over 8}
\right)=g^{2\over 3}~0.599745 \,, \qquad E_{osc}(g) \to g^{2\over 3}
~0.531248 \,, \\
E(g) &\!\!\to\!\!& g^{2\over 3}~0.530181 \,.
\end{eqnarray*}

We observe that our leading term $E_0(g)$ provides result about 15
percent worse than $E_{osc}(g)$, but our present technique with only
parameter $a$ is much easier than the oscillator approximation with
two variational parameters and complicated transformations.

\subsection{Next-to-leading Correction}

Unfortunately, we cannot calculate explicitly the last path integral
in (\ref{hehe1}). On the other hand, we have already factorized out
the main (generalized Gaussian) contribution $e^{-V E_0(g)}$ and
therefore, the remaining part $e^{-V \triangle E(g)}$ should not
result in a relatively strong correction. Therefore, we develop a
systematic scheme to estimate the non-Gaussian correction.

We define the next-to-leading non-Gaussian term as follows:
\begin{eqnarray*}
\triangle E_1(g)
= {1\over 2V}\int d\Sigma_{\cal D}~{\rm Tr}
\left( \ln[1+2ig\phi D]-2g^2 [\phi D\phi D] \right) \,.
\end{eqnarray*}

Taking into account the symmetry $\phi \leftrightarrow -\phi$ we
rewrite
\begin{eqnarray*}
\triangle E_1(g)
= {1\over 4V}\int d\Sigma_{\cal D}~{\rm Tr} \left(
\ln[1+4g^2\phi D \phi D]-4g^2 [\phi D\phi D] \right) \,.
\end{eqnarray*}

Then, going to a new scale
$$
t \rightarrow {1\over \sqrt{a}}~t \,, \quad
k \rightarrow \sqrt{a}~k \,, \quad
\tilde{D}(k) \rightarrow {1\over a(k^2+1+1/a)}  \,, \quad
\tilde{\cal D}(k) \rightarrow 1-{2\over k^2+6+4/a}
$$
we rewrite
\begin{eqnarray*}
\triangle E_1(g)
= {g^2\over 4Va}~{\rm Tr} \int d\Sigma_{\cal D}~
(\ln[1+\Theta]-\Theta)  \,, \qquad \Theta=4\phi D\phi D \,.
\end{eqnarray*}

In general, any direct evaluation of $\triangle E_1(g)$ represents
a heavy and lengthy procedure. But we can easily estimate it as
follows. First, we note that function $\ln(1+x)-x$ is concave while
$\ln(1+x)-x+x^2/2$ is a convex one. Then, by using these properties,
we can easily find a lower and an upper bound to $\triangle E_1(g)$
as follows:
\begin{eqnarray*}
&& \triangle E_1^{-}(g)\le\triangle E_1(g)\le\triangle E_1^{+}(g)\,,\\
&& \triangle E_1^{-}(g)={g^2\over 4Va}~\left\{ \ln[1+\left\langle
\Theta\right\rangle]-\left\langle\Theta\right\rangle-{\left\langle
\Theta^2\right\rangle-\left\langle\Theta\right\rangle^2 \over 2}
\right\} \,,\\
&& \triangle E_1^{+}(g)={g^2\over 4Va}~\left\{ \ln[1+\left\langle
\Theta\right\rangle]-\left\langle\Theta\right\rangle  \right\} \,,
\qquad \left\langle (...) \right\rangle = {1\over V}\int\!
d\Sigma_{\cal D} {\rm Tr}(...) \,.
\end{eqnarray*}

We calculate
\begin{eqnarray*}
&& \left\langle \Theta \right\rangle = 4\int\!{dp\over 2\pi}
\left(1-{2\over p^2+6+4/a}\right) \int\!{dk\over 2\pi}
{1\over k^2+1+1/a} {1\over (k+p)^2+1+1/a}= 4Q(a) \,,\\
&& Q(a)=a{2\sqrt{1+a}+\sqrt{4+6a} \over
4(1+a)^{1/2}(2+3a+\sqrt{(1+a)(4+6a)})} \,,\\
&& \left\langle \Theta^2 \right\rangle = 16~[F(a)+2R(a)] \,,\\
&& F(a)=a^{3/2}{\sqrt{1+a}~(24+31a)-\sqrt{4+6a}~(12+11a) \over
16(1+a)(2+3a)(3+5a)} \,,\\
&& R(a)=a^{3/2}{9\sqrt{(1+a)(4+6a)}-4-6a
\over 32(1+a)^{3/2}(2+3a)(3+5a)} \,.
\end{eqnarray*}

For strong-coupling regime $g\gg 1$ we obtain
$$
\left\langle\Theta\right\rangle = \sqrt{2\over 3}=0.816497\,, \qquad
\left\langle \Theta^2 \right\rangle = {5\over 3}-{2\sqrt{6}\over 115}
=1.62407
$$
so
\begin{eqnarray*}
-g^{2/3}~0.139072\le\triangle E_1(g) \le -g^{2/3}~0.054897
\end{eqnarray*}
and
\begin{eqnarray*}
g^{2/3}~0.460673\le  E_1(g) \le g^{2/3}~0.544848
\end{eqnarray*}
while
$$
E_0(g)=g^{2/3}\cdot 0.599745 \,, \qquad E(g)=g^{2/3}\cdot 0.530181 \,.
$$

We see that the next-to-leading correction $\triangle E_1(g)$
is negative and it may lower the energy $E_0(g)$ about 10 percent
approaching close to the exact numerical result.

\section{Scalar field model: $d=2$}

The scalar $\varphi^4$ theory in two spatial dimensions has been
intensively investigated \cite{simon73,glimm75,mcbr76} as a simple,
but nontrivial example, within which the vacuum exhibits a nontrivial
structure \cite{chang76,drell76,polle88}.

Within the framework of constructive QFT
\cite{simon74,simon73,glimm81,glimm75,rosen76} a set of general
theorems has been proven to establish the existence of nontrivial
two-dimensional theories of self-coupling scalar field. Unfortunately
constructive quantum field theory gave no effective instrument (like
Feynman diagrams) for the calculation of important physical
characteristics of the QFT models.

An attractive approach to the problem under discussion is the
variational method of the Gaussian effective potential
\cite{barne80,barde83,steve84,conso85}. Original investigations in
the same direction were made in \cite{chang75,magru76,baym77,grass91}.
An attempt to go beyond the Gaussian approximation has been made
in \cite{polli89}.

However, specific features of the variational approach in QFT make
their results unreliable if the theory has divergencies (for $d>1$)
in the highest perturbation orders (see, e.g. \cite{feyn88}).

In this paper we calculate the ground-state (vacuum) energy.
We demonstrate an effective scheme resulting in the explicit
and exact asymptotics of the self energy for strong coupling.

Consider a superrenormalizable scalar model $\phi_2^4$ in
two-dimensional Euclidean space-time $x\in V\subset{\bf R}^2$. The
Lagrangian is given:
\begin{eqnarray}
\label{lagr2} &&
 {\cal L}={1\over2}(\phi[\Box-m^2]\phi)-{g^2\over2}:\phi^4:\,,
\end{eqnarray}
The kernel $\Box-m^2$ and its Green function are
\begin{eqnarray*}
&& D^{-1}(x-x')=(-\Box+m^2)\delta(x-x'), \\ &&
D(x)=\int{dk\over(2\pi)^2}~{e^{ikx}\over k^2+m^2}
=\int\limits_0^\infty{d\beta\over 4\pi\beta}~ e^{-{\beta\over
2}m^2-{x^2\over 2\beta}}, \qquad
D_0 = D(0)= \int\limits_0^\infty{d\beta\over 4\pi\beta}~
e^{-{\beta\over 2}m^2} \,.
\end{eqnarray*}

The super-renormalizable theory $\phi^4_2$ contains divergences which
can be removed by the renormalization of mass and vacuum energy.
However both these divergences can be removed if the interaction
Lagrangian in (\ref{lagr2}) is written in the normal form:
\begin{eqnarray}
\label{lagr2n}
:\phi^4:=\phi^4-6D_0\phi^2+3D_0^2=(\phi^2-3D_0)^2-6D_0^2.
\end{eqnarray}
This form of interaction removes all the divergences in this theory.
For intermediate calculations we shall use the dimensional
regularization, i.e. we consider the theory in the space ${\bf R}^d$
($d<2$)
\begin{eqnarray*}
&& D_{{\rm reg}}(x)=\int{d^dk\over(2\pi)^d}~{e^{ikx}\over k^2+m^2}
={1\over 2}
\int\limits_0^\infty{d\beta\over (2\pi\beta)^{d/2}}~
e^{-{\beta\over 2}m^2-{x^2\over 2\beta}}, \\
&& D_0= D_{{\rm reg}}(0)= {1\over 2} \int\limits_0^\infty
{d\beta\over (2\pi\beta)^{d/2}}~e^{-{\beta\over 2}m^2}
={\Gamma\left(1-{d\over2}\right)\over
2(2\pi)^{d/2}}\cdot\left({2\over m^2}\right)^{1-{d\over2}} \,.
\end{eqnarray*}
All divergences are explicitly removed in final formulae and then,
we put the true spatial dimension $d=2$. For simplicity we omit the
index "reg" throughout the text having in mind that the regularization
has been introduced.

It is convenient to rewrite the Lagrangian as follows:
\begin{eqnarray}
{\cal L}={1\over2}(\phi[\Box-m^2]\phi)-
{g^2\over 2}\left(\phi^2-3D_0\right)^2+3g^2D_0^2 \,.
\end{eqnarray}

In this paper we shall calculate the vacuum energy. Let us
consider the partition function and use the Gaussian
representation:
\begin{eqnarray}
\label{ZZ}
 Z_V[g]&=& \det(-\Box+m^2)
\int\delta\phi~e^{-{1\over2}(\phi[-\Box+m^2]\phi)
-{g^2\over2}\left(\left[\phi^2-3D_0\right]^2\right)
+3g^2D_0^2V}                                           \nonumber\\
&=&\int D\Phi ~\det(-\Box+m^2)
\int\delta\phi~e^{-{1\over2}(\phi[-\Box+m^2]\phi)
-ig\left(\Phi\left[\phi^2-3D_0\right]\right)+3g^2D_0^2V}  \nonumber\\
&=&\int D\Phi ~e^{3igD_0(\Phi)
-{1\over2}{\rm Tr}\ln(1+i2g\Phi D)+3g^2D_0^2V} \,,       \nonumber\\
&=&\int D\Phi ~e^{igD_0(\Phi)
-{1\over2}{\rm Tr}\ln(1-4g^2D_0D+i2g\Phi D)-g^2D_0^2V} \,, \qquad
D\Phi= \delta\Phi~e^{-{1\over 2}(\Phi\Phi)}
\end{eqnarray}
with normalization $Z_V[0]=1$.

Let us consider the interaction part
\begin{eqnarray*}
U[\Phi] &&= {\rm Tr}\ln\left[1-4g^2D_0D+i2g\Phi D \right] ={\rm
Tr}\ln\left[ {D^{-1}-4g^2D_0+i2g\Phi\over D^{-1}} \right] \\
&&= \int dx\int\limits_0^\infty{d\alpha\over \alpha}\left.\left[
e^{-{\alpha\over2}(-\Box+m^2)}-e^{-{\alpha\over2}
(-\Box+m^2-4g^2D_0+i2g\Phi(x))}\right]\delta^d(x-x')\right
\vert_{x=x'}      \\
&& =\int dx\left[W_{m^2}(x,x\vert0)
- W_{m^2-4g^2D_0}(x,x\vert\Phi)\right]\,.
\end{eqnarray*}
where the function $W$ is defined as:
\begin{eqnarray*}
W_{m^2}(x,x'\vert\Phi)=\int\limits_0^\infty {d\alpha\over \alpha}
~e^{-{\alpha\over2}(-\Box+m^2+i2g\Phi(x))} \cdot\delta^d(x-x').
\end{eqnarray*}
It satisfies the condition
\begin{eqnarray*}
W_{m^2}(x,x'\vert\Phi+igA)=W_{m^2-2g^2A}(x,x'\vert\Phi).
\end{eqnarray*}
The function $W$ can be represented in the form of a functional
integral (see Appendix A):
\begin{eqnarray}
\label{green1}
W_{m^2}(x,x'\vert\Phi)=\int\limits_0^\infty{d\alpha\over(2\pi)^{d/2}
\alpha^{1+{d\over2}}} e^{-{\alpha
m^2\over2}-{(x-x')^2\over2\alpha}}~ \int d\sigma[\xi]
~e^{-ig\int\limits_0^\alpha\Phi\left(x{\beta\over\alpha}
+x'\left(1-{\beta\over\alpha}\right)+\xi(\beta)\right)} \,,
\end{eqnarray}
where $\xi(\beta)\in{\bf R}^d$ and
\begin{eqnarray*}
d\sigma[\xi]=\delta\xi\cdot e^{-{1\over2}\int\limits_0^\alpha
d\beta~\dot{\xi}^2(\beta)}=\delta\xi\cdot e^{-{1\over2}
(\xi K^{-1}\xi)}, \qquad \int d\sigma[\xi]=1 \,.
\end{eqnarray*}
with the boundary conditions $\xi(0)=\xi(\alpha)=0$.

The Green function of the kernel
$$
K^{-1}(\beta,\beta')=-{d^2\over d\beta^2}\delta(\beta-\beta')
$$
reads
$$
K(\beta,\beta')= -{1\over2}\vert\beta-\beta'\vert
+{1\over2}(\beta+\beta')-{\beta\beta'\over\alpha}
$$ and satisfies
the boundary conditions
$$
K(\beta,\beta')=K(\beta',\beta),~~~~~K(\beta,0)=K(\beta,\alpha)=0.
$$

Then,
\begin{eqnarray}
\label{green0} W_{m^2}(x,x\vert\Phi)
=\int\!\! D\alpha \int d\sigma[\xi] ~e^{-ig(\Phi B_\xi)} \,, \qquad
\int D\alpha \cdot (...) = \int\limits_0^\infty
{d\alpha\over (2\pi\alpha)^{d/2} \alpha}~e^{-{\alpha m^2\over2}}
\cdot (...)\,.
\end{eqnarray}
where
 $$ B_\xi(y)=\int\limits_0^\alpha
d\beta~\delta(y-x-\xi(\beta)), ~~~~~~~\int dy~B_\xi(y)=\alpha \,.
$$ so that
 $$ \int\limits_0^\alpha d\beta~
\Phi\left(x+\xi(\beta)\right) = \int dy~\Phi(y)B_\xi(y)
=(\Phi B_\xi) \,, \quad \int\!\! dx (\Phi B_\xi) = \alpha (\Phi) \,.
$$

Finally we get
\begin{eqnarray*}
U[\Phi]&=& \int\!\! dx \int\!\! D\alpha \int\!\! d\sigma[\xi]
~\left[1-e^{2\alpha g^2D_0-ig(\Phi B_\xi)}\right] \,.
\end{eqnarray*}

Following the idea mentioned in the previous sections, we go to a
shift $$ \Phi(x) \rightarrow \Phi(x)-igA $$ where $A$ is a constant
depending on $g$.

We rewrite
\begin{eqnarray*}
Z_V[g]= e^{g^2\left[{1\over2}A^2+D_0A-D_0^2\right]V}
~\int D\Phi~e^{ig(D_0+A)(\Phi)-{1\over2}U\left[\Phi-igA\right]}
\end{eqnarray*}
and
\begin{eqnarray}
\label{U2} && U\left[\Phi-igA\right]=\int
dx\int\limits_0^\infty{d\alpha\over(2\pi)^{d/2}\alpha^{1+{d\over2}}}
e^{-{\alpha m^2\over2}}\cdot\int d\sigma[\xi]\cdot \left[1-
e^{g^2\alpha (2D_0-A)-ig(\Phi B_\xi)}\right]
\end{eqnarray}

Now we introduce the normal-ordered form:
\begin{eqnarray*}
&& e^{-ig(\Phi B_\xi)} = :e^{-ig(\Phi B_\xi)}:_\Phi
e^{-{g^2\over2}(B_\xi B_\xi)} \,, \\
 && e^{ik(\xi(\beta_1)-\xi(\beta_2))}
=:e^{ik(\xi(\beta_1)-\xi(\beta_2))}:_\xi~e^{-{k^2\over 2}
F_{\alpha}(|\beta_1-\beta_2|)}\,, \qquad
F_{\alpha}(|\beta_1-\beta_2|)=
|\beta_1-\beta_2|- {(\beta_1-\beta_2)^2\over\alpha}
\end{eqnarray*}
obeying the following relations
\begin{eqnarray*}
\int D\Phi~ :e^{-ig(\Phi B_\xi)}:_\Phi =1,~~~~~~
\int d\sigma[\xi]:e^{ik(\xi(\beta_1)-\xi(\beta_2))}:_\xi =1 \,.
\end{eqnarray*}

One gets
\begin{eqnarray*}
(B_\xi B_\xi)&=&\int dy B_\xi(y)B_\xi(y)=
 \int\!\!\!\!\int\limits_0^\alpha d\beta_1
d\beta_2 ~\delta^d(\xi(\beta_1)-\xi(\beta_2))\\
&=&\int\!\!\!\!\int\limits_0^\alpha d\beta_1 d\beta_2
\int{d^dk\over(2\pi)^d}:e^{ik(\xi(\beta_1)-\xi(\beta_2))}:_\xi
e^{-{k^2\over2}F_{\alpha}(|\beta_1-\beta_2|)}
=\left\langle(B_\xi B_\xi)\right\rangle_\xi +:W_\alpha[\xi]:_\xi \,,
\end{eqnarray*}
where
\begin{eqnarray*}
\left\langle(B_\xi B_\xi)\right\rangle_\xi
&& = \int d\sigma[\xi](B_\xi B_\xi)
=\int\!\!\!\!\int\limits_0^\alpha d\beta_1 d\beta_2
\int{d^dk\over(2\pi)^d}e^{-{k^2\over2}
F_{\alpha}(|\beta_1-\beta_2|)} \\
&& = {2\over(2\pi)^{d/2}}\int\limits_0^\alpha d\beta
(\alpha-\beta)^{1-{d\over2}}
\left({\alpha\over\beta}\right)^{{d\over2}}
={2\alpha^{2-{d\over2}}\over(2\pi)^{d/2}}
B\left(2-{d\over2},1-{d\over2}\right)
\end{eqnarray*}
and the interaction functional is introduced
\begin{eqnarray*}
:W_\alpha[\xi]:_\xi &=& \int\!\!\!\!\int\limits_0^\alpha
d\beta_1 d\beta_2 \int{d^dk\over(2\pi)^{d}}
:e_0^{ik(\xi(\beta_1)-\xi(\beta_2))}:_\sigma
e^{-{k^2\over2}F_{\alpha}(|\beta_1-\beta_2|)} \,,             \\
e^z_k &=& e^z-\sum\limits_{s=0}^k{z^s\over s!}\,, \qquad \int
d\sigma[\xi]~:W_\alpha[\xi]:_\xi=0 \,.
\end{eqnarray*}
It is easy to check that
\begin{eqnarray*}
e^{g^2\alpha(2D_0-A)-ig(\Phi B_\xi)}
&& =e^{-g^2\alpha A +2g^2\alpha D_0-{g^2\over2} (B_\xi B_\xi)}
\cdot :e^{-ig(\Phi B_\xi)}:_\Phi                          \\
&& =e^{-g^2\alpha A-g^2N(\alpha)}\cdot
e^{-{g^2\over2}:W_\alpha[\xi]:_\xi}\cdot \left[
:e_1^{-ig(\Phi B_\xi)}:_\Phi+1-ig(\Phi B_\xi)\right]  \,,
\end{eqnarray*}
where
\begin{eqnarray*}
-g^2N(\alpha)
&& =2\alpha g^2D_0- {g^2\over2} \left\langle(B_\xi B_\xi)
\right\rangle_\xi     \\
&& =g^2\left\{2\alpha\cdot{\Gamma\left(1-{d\over2}\right)\over
2(2\pi)^{d/2}}\left({2\over m^2}\right)^{1-{d\over2}}-
{1\over2}\cdot{2\alpha^{2-{d\over2}}\over(2\pi)^{d/2}}
\cdot{\Gamma\left(2-{d\over2}\right)\Gamma\left(1-{d\over2}\right)
\over \Gamma\left(3-d\right)}\right\}     \\
&& ={g^2\alpha\over(2\pi)^{d/2}}\Gamma\left(1-{d\over2}\right)
\left({2\over m^2}\right)^{1-{d\over2}}
\left\{1-\left({m^2\alpha\over2}\right)^{1-{d\over2}}
\cdot{\Gamma\left(2-{d\over2}\right)\over
\Gamma\left(3-d\right)}\right\}= (d\to 2) \\
&& =-{\alpha g^2\over2\pi}\left[{\bf C}+ \ln\left({\alpha
m^2\over2}\right)\right] \,, \qquad {\bf C}=0.577215...
\end{eqnarray*}

Then, the functional $U$ in (\ref{U2}) can be represented in the form
\begin{eqnarray*}
U\left[\Phi-igA\right] \!\!\!\! &&=\int\!\! dx \int\!\! D\alpha
\int d\sigma[\xi]~\left[1-e^{Q(g^2\alpha,~A)}
~e^{-{g^2\over2}:W_\alpha[\xi]:_\xi}\cdot\left[
:e_1^{-ig(\Phi B_\xi)}:_\Phi+1-ig(\Phi B_\xi)\right]\right] \\
&& = VU_0+ig(\Phi)U_1+U_I[\Phi]     \,,
\end{eqnarray*}
where
\begin{eqnarray*}
Q(g^2\alpha,~A) &=& -g^2N(\alpha)-g^2\alpha A \,, \qquad \qquad
R\left(g^2\alpha\right)=\int d\sigma[\xi]
~e^{-{g^2\over 2}:W_\alpha[\xi]:_\xi} \,, \\
U_0&=& \int\!\! D\alpha \left[1- e^{Q(g^2\alpha,~A)}
R\left(g^2\alpha\right)\right] \,, \qquad
U_1=\int\!\!  D\alpha ~\alpha~e^{Q(g^2\alpha,~A)}
R\left(g^2\alpha\right)\,,\\
U_I[\Phi]&=&-\int\!\! dx \int D\alpha ~e^{Q(g^2\alpha,~A)}
\int d\sigma[\xi] ~e^{-{g^2\over 2}:W_\alpha[\xi]:_\xi}
~:e_1^{-ig(\Phi B_\xi)}:_\Phi \,.
\end{eqnarray*}

The optimal value of parameter $A$ should be obtained from the
condition of elimination of the total linear terms over $\Phi$:
\begin{eqnarray*}
ig~(\Phi)\left[D_0+A-{1\over 2}U_1\right]=0 \,,
\end{eqnarray*}
i.e. this equation reads
\begin{eqnarray*}
&& A=-{1\over 2}\int D\alpha ~\alpha \left[1-e^{Q(g^2\alpha,~A)}
R\left(g^2\alpha\right)\right] \,.
\end{eqnarray*}

The vacuum energy in the lowest approximation looks
\begin{eqnarray*}
E_0&=&-g^2 \left({1\over2}A^2+D_0A-D_0^2\right)
+ {1\over2}U_0=-{g^2\over2}A^2+W_0 \,,  \\
W_0&=& {1\over 2} \int D\alpha \left\{ e^{Q(g^2\alpha,~A)}
-1-Q(g^2\alpha,~A)+e^{Q(g^2\alpha,~A)} [1-R(g^2,\alpha)] \right\}\,,
\end{eqnarray*}
where divergences are completely removed as follows:
\begin{eqnarray*}
-g^2 D_0 A+g^2 D_0^2-{1\over 2}\int\!\! D\alpha~Q(g^2\alpha,~A)=0 \,.
\end{eqnarray*}

After complete removal of divergences we can put $d=2$. Then,
\begin{eqnarray*}
A&=&-\int\limits_0^\infty {d\alpha\over 4\pi\alpha}
~e^{-{\alpha m^2\over2}}\left[1-e^{Q(g^2\alpha,~A)}
R\left(g^2,\alpha\right)\right]                         \\
W_0&=&\int\limits_0^\infty{d\alpha\over 4\pi\alpha^2}
~e^{-{\alpha m^2\over2}} \left[e^{Q(g^2\alpha,~A)}-1
-Q(g^2\alpha,~A)+e^{Q(g^2\alpha,~A)}\cdot (1-R(g^2\alpha)) \right] \,.
\end{eqnarray*}

Going to new re-scaled variables:
$$
\beta=\alpha\tau,~~~~\xi(\beta)=\sqrt{\alpha}~\eta(\tau),~~~~
k={q\over\sqrt{\alpha}},~~~~~~s={g^2\alpha\over2}=ht \,, \qquad
B={A\over h}
$$
with
$$ N(t)=t\left[{\bf C}+\ln(t)\right] \,, \qquad
\alpha={2\over m^2}~t \,, \qquad h={g^2\over\pi m^2} \,, \qquad
$$
we obtain
\begin{eqnarray}
Z[g]=e^{-VE}=e^{-VE_0} \int D\Phi ~e^{-{1\over 2}U_I[\Phi]}
= e^{-VE_0}~e^{-VE_{corr}}\,,
\label{Zfin}
\end{eqnarray}
where the leading-order term for the vacuum energy is
\begin{eqnarray}
\left({m^2 \over 8\pi}\right)^{-1} \!\! E_0 = {\cal E}_0
= -h \left\{B^2 + \int\limits_0^\infty {dt\over t}~e^{-t}~{1\over ht}
\left[ e^{Q(ht)}-1-Q(ht)+e^{Q(ht)}~ \left(R(ht)-1 \right)~\right]
\right\}
\label{Efin}
\end{eqnarray}
and the remaining higher-order correction reads
\begin{eqnarray}
E_{corr}=-{1\over V} \ln \int D\Phi ~e^{-{1\over 2}U_I[\Phi]} \,.
\label{Ecorr}
\end{eqnarray}

The shift parameter $B$ is governed by the equation:
\begin{eqnarray}
B={1\over 2}\int\limits_0^\infty{dt\over t}
~e^{-t}\left[e^{Q(ht)}~R(ht)-1~\right] \,,
\label{Bfin}
\end{eqnarray}
where
\begin{eqnarray*}
&& Q(ht)=-h~t~[B+{\bf C}+\ln(t)] \,,  \qquad
   R(ht)=\int d\sigma[\eta]~e^{-ht :W[\eta]:_{\sigma} } \,, \\
&& d\sigma[\eta]=\delta\eta\cdot e^{-{1\over2}\int\limits_0^1
   d\tau~\dot{\eta}^2(\tau)}, \qquad
   \int d\sigma[\eta]~:W[\eta]:_\sigma =0 \,, \\
&& :W[\eta]:_\sigma = \int\!\!\!\!\int\limits_0^1
   d\tau_1 d\tau_2 \int{dq\over(2\pi)^2}
   :e_0^{iq(\eta(\tau_1)-\eta(\tau_2))}:_\sigma
   ~e^{-{q^2\over2}\left(|\tau_1-\tau_2| -
   (\tau_1-\tau_2)^2\right)} \,.
\end{eqnarray*}

The Green function for the kernel of $d\sigma[\eta]$ is
$$
K(t,s)=-{|t-s| \over 2} + {t+s \over 2} -ts \,.
$$

Eqs. (\ref{Zfin}), (\ref{Efin}), (\ref{Ecorr}) and (\ref{Bfin}) are
basic in our consideration and they completely define the ground-state
energy of the system at any given coupling. For general coupling these
equations should be evaluated by numerical means, but we are able to
solve them explicitly in the weak- and strong-coupling regimes.

\subsection{Explicit weak-coupling solutions}

{\bf i}. First, we calculate the exact perturbation solution up to
the order $O(g^6)$ as follows:
\begin{eqnarray*}
e^{-V E(g)}
&&= 1+{1\over 2}~{g^4\over 4}~4!~V \int\!\! d^2x D^4(x)+O(g^6) \\
&&= 1+{3g^4~V \over (2\pi)^3 m^2} \int\limits_{0}^{\infty} \!\!
du~u~K_0^4(u)+O(g^6)
= 1+{3g^4~V \over (2\pi)^3 m^2}~{7\over 8}\zeta(3) +O(g^6) \,, \\
E &&= - {g^4 \over (2\pi)^3 m^2}~{21\over 8}~\zeta(3) + O(g^6)
= - h^2 ~{m^2 \over 8\pi}~ {21\over 8}~\zeta(3) + O(h^3) \,,
\label{Epertur}
\end{eqnarray*}
where $K_0(u)$ and $\zeta(x)$ are the Bessel and the Riemann
function, correspondingly.

{\bf ii}. Now we evaluate $E_{corr}(g)$. We obtain the explicit
weak-coupling solution to the higher-order total correction as
follows (see Appendix C):
\begin{eqnarray}
E_{corr}(g)=-{g^4 \over (2\pi)^3 m^2}~{7\over 8}~\zeta(3) + O(g^6)
= - h^2 ~{m^2 \over 8\pi}~ {7\over 8}~\zeta(3) + O(h^3) \,.
\label{Ecorrweak}
\end{eqnarray}

{\bf iii}. We have $E=E_0+E_{corr}$ by the very definition
(\ref{Zfin}). Therefore, the explicit weak-coupling solution to the
leading-order energy reads:
\begin{eqnarray*}
E_0 = - {g^4 \over (2\pi)^3 m^2}~{7\over 4}~\zeta(3) + O(g^6)
= - h^2 ~{m^2 \over 8\pi}~ {7\over 4}~\zeta(3) + O(h^3) \,.
\label{E_0}
\end{eqnarray*}

Therefore, the following relations take place (within the given
accuracy)
\begin{eqnarray*}
E_{corr}(g) = {1\over 2}~E_0(g) = {1\over 3}~E(g) \,.
\end{eqnarray*}

One can note that our leading-order energy $E_0$ underestimates
(i.e., $2/3$ of) the exact energy $E$. The gap of one third is
compensated by taking into account the higher-order energy term
(\ref{Ecorrweak}). As we will show below, our leading-order
approximation becomes better as $g$ grows and, approaches the
exact solution for $g\to\infty$.

\subsection{Exact strong-coupling solutions}

However, our method is designed to investigate the strong-coupling
behaviour of the considered system. Below we demonstrate explicit
and analytic solutions of Eqs. (\ref{Zfin}), (\ref{Efin}) and
(\ref{Bfin}) that leads to the exact strong-coupling answer.

We note that for $h\to\infty$ the factor $R(ht)$ gives correction
$\sim e^{O(\ln(h)/h)}$ with respect to the main exponentials in
(\ref{Efin}) and (\ref{Bfin}). Therefore, one can neglect it
considering the main asymptotical values for $B$ and $E_0$
(for more details see Appendix B).

For $h\to\infty$ the integrand in (\ref{Bfin}) behaves a very sharply
expressed Gaussian exponential centered at point $t_m$. The exponent
function is
$$
f(t)=-t-\ln(t)-ht[B+{\bf C}+\ln(t)] \,.
$$
Then, with the use of the "saddle-point" method we estimate
\begin{eqnarray}
B = {1\over 2} \int\limits_{-\infty}^{\infty} dt
\exp\left\{ f(t_m)-{1\over 2}~t^2 [-f''(t_m)] \right\}
= {1\over 2} \exp\left\{ f(t_m) \right\} \sqrt{2\pi\over -f''(t_m)}\,,
\end{eqnarray}
where the maximum point $t_m$ obeys the conditions
$$
f'(t_m)=0\,, \qquad f''(t_m)<0 \,.
$$

By resolving these equations we obtain (see Appendix D):
\begin{eqnarray}
B=\ln(h)-{\bf C}-2+O(1/\ln^4(h)) \,.
\label{asympB}
\end{eqnarray}
Note, this asymptotics is reached very slowly, e.g., for $h=10^{20}$
we obtain $B=0.93265\cdot\ln(h)$ instead of $B=\ln(h)$. Substituting
(\ref{asympB}) into (\ref{Efin}) and by neglecting terms vanishing as
$O(1/h)$ we obtain (Appendix D):
\begin{eqnarray}
{\cal E}_0 = -{3\over 2}h\ln^2(h)+3({\bf C}+2)h\ln(h)+O(h) \,.
\label{asympE}
\end{eqnarray}
Particularly, for $h=10^{20}$ this asymptotics provides
${\cal E}_0=-2.82\cdot 10^{23}$ while the numerical result is
${\cal E}_0=-2.81\cdot 10^{23}$. We see that this asymptotical
behaviour becomes exact for very large coupling.

\section{Conclusion}

We have represented a path-integral technique suitable to evaluate
the quartic self interaction in the strong-coupling regime. The
leading-order approximation is obtained in a relatively simple way
and the remaining corrections can be estimated. A simple version of
our method has been tested in the examples of a plain
quartic-exponential function and the well-investigated anharmonic
oscillator in Quantum Mechanics. We see that our technique effectively
isolates the main contribution for arbitrary coupling. Then, we have
applied this technique to the solution of the ground-state energy in
the superrenormalizable scalar theory $\phi^4(x)$ in two dimensions.
Hereby, the auxiliary appearing complex functional does not represent
any difficulties for this technique. Our leading-order approximation
becomes much accurate as the coupling $g$ increases and for $g\to\infty$
it coincides with the exact asymptotics. We can conclude that our
technique may effectively isolate the main contribution of the
strong-coupling regime even in theories with divergencies and complex
functionals.

\begin{table}[ht]
\begin{center}
\begin{minipage}{12cm}
{\bf Table 1.}
{\sl
Approximate and exact numerical results for $E(g)$ and $G(g)$
in zero dimension.}
\end{minipage}

\vspace*{5mm}
{\small
\begin{tabular}{|c|c|c|c|c|c|c|}
\hline
        &        &        &      &        &        &      \\
$g^2/2$ &$E_0(g)$&$E_1(g)$&$E(g)$&$G_0(g)$&$G_1(g)$&$G(g)$ \\
        &        &        &      &        &        &      \\
\hline
0.01 & 0.02783 & 0.02626 & 0.02629 & 0.90518 & 0.90650 & 0.90653  \\
0.1  & 0.18027 & 0.15267 & 0.15361 & 0.60786 & 0.61378 & 0.61553  \\
0.2  & 0.27274 & 0.22892 & 0.23000 & 0.49492 & 0.50023 & 0.50312  \\
0.5  & 0.42431 & 0.35925 & 0.35993 & 0.35854 & 0.36211 & 0.36596  \\
1.0  & 0.55590 & 0.47747 & 0.47758 & 0.27256 & 0.27489 & 0.27884  \\
2.0  & 0.69826 & 0.60939 & 0.60890 & 0.20316 & 0.20461 & 0.20823  \\
5.0  & 0.89861 & 0.79988 & 0.79874 & 0.13478 & 0.13553 & 0.13840  \\
10.  & 1.05688 & 0.95301 & 0.95150 & 0.09766 & 0.09811 & 0.10039  \\
100. & 1.60679 & 1.49423 & 1.49209 & 0.03217 & 0.03226 & 0.03313  \\
\hline
\end{tabular}}
\end{center}
\end{table}

\begin{table}[ht]
\begin{center}
\begin{minipage}{10cm}
{\bf Table 2.}
{\sl
Comparison of results for $E_0(g)$ , $E_{osc}(g)$ and $E_{num}(g)$ in
one dimension.
}
\end{minipage}

\vspace*{5mm}
{\small
\begin{tabular}{|c|c|c|c|}
\hline
       &            &            &   \\
${g^2\over 2}$  &  $E_0(g)$  & $E_{osc}(g)$ & $E_{num}(g)$ \\
       &            &            &   \\
\hline
  0.1  &   0.06434  &   0.05938  &   0.05915   \\
  0.5  &   0.22666  &   0.19697  &   0.19618   \\
  1.0  &   0.35522  &   0.30490  &   0.30377   \\
 10.0  &   1.17291  &   1.00778  &   1.00497   \\
 50.0  &   2.30990  &   2.00461  &   1.99971   \\
100.0  &   3.02802  &   2.63759  &   2.63138   \\
\hline
\end{tabular}}
\end{center}
\end{table}

\vskip 1cm
{\Large {\bf Appendix A. }}
\vspace*{5mm}

The Green function in the presence of an external field reads:
\begin{eqnarray*}
G_m(t,t'|\Phi)&=&{1\over D^{-1}+m^2+2ig\Phi(x)}\delta(t-t')\\
&=&\int\limits_0^\infty ds~e^{-s(1+a+m^2)}\cdot{\rm T}~
e^{\int\limits_0^sd\tau\left({d\over dt(\tau)}\right)^2
-2ig\int\limits_0^sd\tau~\Phi(t(\tau))}\delta(t-t') \,.
\end{eqnarray*}

Then,
\begin{eqnarray*}
&&{\rm T}~e^{\int\limits_0^sd\tau\left({d\over dt(\tau)}\right)^2
-2ig\int\limits_0^sd\tau~\Phi(t(\tau))}\delta(t-t') \\
&&=\int D\nu~e^{-\int\limits_0^sd\tau~\nu^2(\tau)+
2\int\limits_0^sd\tau\nu(t){d\over dt(\tau)}
-2ig\int\limits_0^sd\tau~\Phi(t(\tau))}\delta(t-t') \\
&&=\int D\nu~e^{-\int\limits_0^sd\tau~\nu^2(\tau)
-2ig\int\limits_0^sd\tau~\Phi\left(t+2\int\limits_t^sd\tau'\nu(\tau')
\right)} \delta\left(t-t'+2\int\limits_0^sd\tau\nu(\tau)\right)  \\
&&=\int D\xi\int\limits_{-\infty}^\infty d\nu_0\sqrt{{s\over\pi}}~
e^{-s\nu_0^2-\int\limits_0^sd\tau~\dot{\xi}^2(\tau)
-2ig\int\limits_0^sd\tau~\Phi(t+2(s-\tau)\nu_0-2\xi(\tau))}
\delta(t-t'+2s\nu_0)\\
&&={1\over\sqrt{4\pi s}}e^{-{(t-t')^2\over4s}}\cdot
\int\limits_{\xi(0)=\xi(s)=0}D\xi~e^{-\int\limits_0^sd\tau
~\dot{\xi}^2(\tau)-2ig\int\limits_0^sd\tau~\Phi\left(t{\tau\over s}
+t'\left(1-{\tau\over s}\right)-2\xi(\tau)\right)} \,,
\end{eqnarray*}
where
$$
\nu(\tau)=\nu_0+\dot{\xi}(\tau),~~~~~~~\xi(0)=\xi(s)=0 \,.
$$

\vskip 1cm
{\Large {\bf Appendix B. }}
\vspace*{5mm}

The following effective approximation takes place:
$$
R(s)=\int d\sigma[\eta]~e^{-sW} = \left\langle ~e^{-sW}
\right\rangle \approx e^{-s\left\langle W \right\rangle} \cdot
\cosh\left( s\sqrt{\left\langle W^2\right\rangle-\left\langle
W\right\rangle^2} \right)\,, \qquad W=:W[\eta]:_\sigma \,.
$$

Due to normal-ordered form of $W$, one finds
$\left\langle W \right\rangle =0$. Then,
$$
R(s) \approx \cosh(sw)\,, \qquad
w^2=\int d\sigma[\eta]~\left( :W[\eta]:_\sigma \right)^2 \,.
$$
For $h\to\infty$ its correction to the shift parameter $B=\ln(h)$
is $\sim O(1)$ and to the leading-order energy $E_0=-(3/2)h\ln^2(h)$
-- $\sim O(h)$. Therefore, its influence is negligible and we just
drop the factor $R(ht)$ by considering the large $h$ asymptotics.

\vskip 1cm
{\Large {\bf Appendix C. }}
\vspace*{5mm}

By expanding the interaction functional
$$
U_I[\Phi] = g^2 U_2[\Phi] + g^4 U_4[\Phi] + O(g^6)
$$
and taking into account its normal-ordered form, we obtain
$$
E_{corr}(g)=-{g^4\over 32 V} \int\!\!\!\!\int\!\! D\alpha D\beta
\int\!\!\!\!\int\!\! d\sigma[\xi] d\sigma[\eta] \int\!\!\!\!\int\!\!
dx dy \int\!\! D\Phi :(\Phi B_{\xi})^2_{x\alpha}:
:(\Phi B_{\eta})^2_{y\beta}: +~O(g^6)\,.
$$
The functional averaging over field $\Phi$ results in
$$
\int\!\! D\Phi :(\Phi B_{\xi})^2_{x\alpha}::(\Phi B_{\eta})^2_{y\beta}:
= 2\int\!\!\!\!\int\limits_{0}^{\alpha} \!\! dtdu
\int\!\!\!\!\int\limits_{0}^{\beta} \!\! dsdw
\int\!\! {d^2k \over (2\pi)^2} \int\!\! {d^2p \over (2\pi)^2}
~e^{i(k+p)(x-y)+ik[\xi(t)-\eta(s)]+ip[\xi(u)-\eta(w)]} \,.
$$
Further we use the following plain relations:
\begin{eqnarray*}
&& \int\!\! dx \int\!\! dy~e^{i(k+p)(x-y)}=V\cdot (2\pi)^2\cdot
\delta^2(k+p) \,, \\
&& \int\!\! d\sigma[\xi] ~e^{ik[\xi(t)-\xi(s)]}
=\exp\left[-{k^2\over 2} F_{\alpha}(|t-s|)\right] \,, \\
&& \int\!\!\!\!\int\limits_{0}^{\alpha} \!\! dt~ds
~e^{-{k^2\over 2} F_{\alpha}(|t-s|)} = 2\alpha
\int\limits_{0}^{\alpha} \!\! dt ~e^{-{k^2\over 2} F_{\alpha}(t)}
~\left( 1-{t\over\alpha} \right) \,, \\
&& \int\!\! {d^2k \over (2\pi)^2}
~e^{-{k^2\over 2}[F_{\alpha}(t) + F_{\beta}(s)]}
= {1\over 4\pi} \int\limits_{0}^{\infty}
\!\! du ~e^{-{u\over 2}[F_{\alpha}(t) + F_{\beta}(s)]} \,.
\end{eqnarray*}

Then,
\begin{eqnarray*}
E_{corr}(g) &&= -{g^4\over 8(2\pi)^4} \int\!\!\!\!\int\limits_{0}^{1}
\!\!dxdy (1-x)(1-y) \int\limits_{0}^{\infty} \!\! du
\int\!\!\!\!\int\limits_{0}^{\infty} \!\!d\alpha d\beta
e^{-{\alpha\over 2}[m^2+ux(1-x)]-{\beta\over 2}[m^2+uy(1-y)]} \\
&&= -{g^4\over 16\pi^3 m^2} \int\!\!\!\!\int\limits_{0}^{1}
\!\!dxdy (1-x)(1-y) {\ln[x(1-x)]-\ln[y(1-y)] \over x(1-x)-y(1-y)} \\
&&= -{g^4 \over (2\pi)^3 m^2}~{7\over 8}~\zeta(3)\,.
\end{eqnarray*}

\vskip 1cm
{\Large {\bf Appendix D. }}
\vspace*{5mm}

The maximum point $t_m$ of the sharp exponential function
$$
f(t)=-t-\ln(t)-ht[B+{\bf C}+\ln(t)]
$$
is dictated by the conditions
$$
f'(t_m)=-1-{1\over t_m}-h[B+{\bf C}+1+\ln(t_m)]=0\,, \qquad
f''(t_m)={1-ht_m\over t_m^2} <0 \,.
$$
This results in
$$
B=\ln(h)-\ln(\xi)-{\bf C}-1-{1\over\xi}-{1\over h}
= \sqrt{\pi\over 2(\xi-1)}~e^{\xi+1} \,, \qquad \xi=ht_m>1 \,.
$$
Hereby,
$$
\lim\limits_{h\to\infty} \xi(h) = +1\,.
$$

The solution is
$$
B=\ln(h)-{\bf C}-2+O(1/\ln^4(h)) \,.
$$

For the energy, we rewrite it in a more convenient form
$$
{\cal E}_0 = -h \left\{ B^2 -\int\limits_0^\infty {dt\over t}
~e^{-t}~[F(t)-\Phi(ht)] \right\}\,,
$$
where
$$
F(t)=\int\limits_t^\infty {ds\over s}~e^{-s}+\ln(t)+{\bf C}
=\sum\limits_{n=1}^{\infty} (-1)^{n+1}{t^n\over n!~n}
$$
and
$$
\Phi(ht)={1\over ht}\left\{
e^{-ht[B+{\bf C}-\ln(h)+\ln(ht)]}-1+ht[B+{\bf C}-\ln(h)+\ln(ht)]
\right\}\,.
$$
Therefore,
$$
\int\limits_0^\infty {dt\over t}~e^{-t}~F(t) = {\pi^2\over 12}
\sim O(1)
$$
and
$$
\int\limits_0^\infty {dt\over t}~e^{-t}~\Phi(ht)
={1\over 2} \ln^2(h) + O(\ln(h)) \,.
$$
Substituting these solutions to the energy, one obtains
$$
{\cal E}_0 = -{3\over 2}h\ln^2(h)+3({\bf C}+2)h\ln(h)+O(h) \,.
$$


\end{document}